# One-Way Hyperbolic Metasurfaces Based on Synthetic Motion

Yarden Mazor[1], and Andrea Alù[1,2,3,4*]

[1]Department of Electrical and Computer Engineering, The University of Texas at Austin, Austin, TX 78712, USA

[2]Photonics Initiative, Advanced Science Research Center, City University of New York, New York, NY 10031, USA

[3]Physics Program, Graduate Center, City University of New York, New York, NY 10026, USA

[4]Department of Electrical Engineering, City College of New York, New York, NY 10031, USA

*Moving metasurfaces support guided waves exhibiting unusual optical properties, including strong anisotropy, nonreciprocity and hyperbolic dispersion. However, for these phenomena to be noticeable, high speeds are typically required, challenging their practical implementation. Here, we show a viable route towards the realization of one-way hyperbolic propagation in metasurfaces placed synthetically in motion through traveling-wave space-time modulation. In addition to non-reciprocal hyperbolic propagation, the proposed time-modulation scheme induces additional exotic opportunities for nanophotonic systems, such as efficient non-reciprocal frequency conversion and mode transfer.*

Hyperbolic dispersion enables the propagation of directional waves with high spatial momentum over broad bandwidths [1], particularly useful for imaging, enhanced light-matter interactions and optical emission [2–7]. This unusual regime is conventionally achieved using metamaterials composed of oppositely signed permittivity multilayers in the case of bulk media [2], or nanostrips for hyperbolic metasurfaces [4,8–10]. In order to realize non-reciprocal hyperbolic transport, leading to hyperbolic propagation with unidirectionality and isolation, Ref. [11] explored the use of a layered metamaterial biased by static magnetic fields. A drawback of this approach is the need for strong static magnetic fields, arising from weak magneto-optic response in natural materials, hindering possible extension of these concepts to optics in a realistic integrated nanophotonic setup. Propagation in moving media also features nonreciprocal transport, extreme anisotropy and hyperbolic dispersion [12–16], and these motion-induced effects can arise also along moving surfaces for highly confined surface waves [17]. One of the major hurdles of this implementation, however, is the typically large velocities required to establish these effects, non-negligible compared to the speed of light.

In a parallel research route, space-time modulated systems have proven to be excellent candidates to implement nonreciprocal phenomena in a magnet-free, integrable platform. Wave propagation in space-time modulated waveguides was explored in [18–20], and many recent efforts have suggested the use of temporal modulation to realize nonreciprocal signal transport. Prominent examples include electrically driven nonreciprocity in waveguides [21,22], nonreciprocal transmission and reception using modulated antennas [23,24], one-way efficient signal transmission [25], and magnetless circulators [26,27]. Dynamic phase modulation has been used to design tunable, nonreciprocal metasurfaces enabling control of the scattered waves [25,28] to realize magnetless nonreciprocal metasurfaces, and in [29] a design implementing these effects with non-linear circuit elements has been introduced.

Here, we use space-time modulation of resonant metasurfaces to mimic mechanical motion, tailoring the dispersion in order to induce nonreciprocal hyperbolic surface wave propagation analogous to moving surfaces at large speeds. In order to motivate and later compare the approaches, we first briefly review wave propagation over a moving surface [17]. Consider a homogeneous impedance surface $Z_s = j\eta_0 \bar{X}_s$ moving at velocity $v$ along $\hat{\mathbf{z}}$, shown in figure (1a). The moving surface response satisfies the boundary condition

$$\hat{\mathbf{x}} \times \Delta \mathbf{H} = Z_s^{-1} diag(\gamma, \gamma^{-1}) \mathbf{E}_{\tan} + Z_s^{-1} \gamma (v\hat{\mathbf{z}} \times \mu_0 \mathbf{H}_x) + (v\hat{\mathbf{z}})\hat{\mathbf{x}} \cdot \varepsilon_0 \Delta \mathbf{E}, \tag{1}$$

where $\Delta \mathbf{H}, \Delta \mathbf{E}$ are the field discontinuities across the boundary and $\gamma = (1 - v^2/c^2)^{-1/2}$. A stationary surface guides transverse-magnetic (TM) surface waves with phase velocity $v_p(\bar{X}_s) = c(1 + 4\bar{X}_s^2)^{-1/2}$ [30]. When $v > v_p$, it supports non-reciprocal hyperbolic dispersion [17], as calculated in figure (1b) for $\bar{X}_s = 4.5$ with normalized velocity $\beta = v/c = 0.114$. Figure (1c) shows the response of the moving surface to a stationary point-source excitation located $\sim 0.02\lambda_0$ above the surface, featuring highly asymmetric plasmon excitation due to nonreciprocity. Combined with the broad bandwidth over which an impedance surface supports confined wave propagation, this response can be of interest for several nanophotonic applications benefiting from strong Purcell enhancement over broad bandwidths combined with one-way features. Yet, given the impractically large velocities required, in the following we explore synthetic motion established by spatio-temporal traveling wave modulation, as a way to provide a robust, tunable platform for the implementation of these exotic wave phenomena.

*Formulation.* A space-time modulated metasurface implemented using modulated variable capacitors and inductors can be generally described by the impedance operator $\hat{Z}_s$ [25]:

$$\mathbf{E}_{\tan}(\mathbf{r}_t,t) = \hat{Z}_s[\mathbf{J}_s(\mathbf{r}_t,t)] = L_s(\mathbf{r}_t,t)\frac{\partial \mathbf{J}_s(\mathbf{r}_t,t)}{\partial t} + C_s^i(\mathbf{r}_t,t)\int^t \mathbf{J}_s(\mathbf{r}_t,t), \quad (2)$$

where $\mathbf{r}_t$ is a point on the surface, $\mathbf{E}_{\tan}$ is the electric field component tangent to the surface, $L_s$ is the surface inductance and $C_s^i$ the inverse capacitance, both assumed to be periodic functions of space and time. In [31], we derive a detailed analytical formalism describing wave interactions over this metasurface. If we assume that the surface impedance experiences negligible temporal dispersion in the relevant frequency band of interest, and that the surface elements are modulated by a cosine function, as shown schematically in figure (2a), the impedance operator can be simplified to the scalar expression [19]

$$Z_s = j\eta_0 \bar{X}_s \left(1 - M \cos\left[\Omega t - \frac{2\pi}{a}z\right]\right), \quad (3)$$

$a$ being the modulation period, $\Omega$ the angular frequency, $M$ the modulation depth, and $v_m = \frac{\Omega a}{2\pi}$ the modulation wave velocity. Unlike a mechanically moving surface, space-time modulation produces frequency mixing, and generates new harmonics. The 2D dispersion of the supported surface waves can be calculated after representing them as an infinite sum of TM Floquet harmonics [20,25]

$$\mathbf{H} = \sum_{n=-\infty}^{\infty} H_n^{1,2} (\cos\varphi_n \hat{\mathbf{z}} - \sin\varphi_n \hat{\mathbf{y}}) e^{\mp\alpha_n x - jk_{ty}y - j(k_{tz}+n\kappa)z} e^{j\omega_n t} + C.C, \quad (4)$$

where 1 (2) refers to above (below) the surface, $\kappa = 2\pi/a$, $k_{tn} = \sqrt{k_{ty}^2 + (k_{tz}+n\kappa)^2}$, $\cos\varphi_n = \frac{k_{tz}+n\kappa}{k_{tn}}$, $\sin\varphi_n = \frac{k_{ty}}{k_{tn}}$, $\omega_n = \omega + n\Omega$, $k_{0n} = \frac{\omega_n}{c}$, and we must satisfy $k_{0n}^2 + \alpha_n^2 = k_{tn}^2$.

*Dispersion and spectral characteristics.* By applying the boundary conditions on the modulated surface, we obtain a recursive relation between the harmonic amplitudes:

$$H^1_{n-1,(y,z)} + \frac{2}{M} D_n H^1_{n,(y,z)} + H^1_{n+1,(y,z)} = 0 \; , \tag{5}$$

with $D_n = 1 - \left(2\bar{X}_s k_{0n} a\right)^{-1} \sqrt{\left(\mathbf{k}_t a + 2\pi n \hat{\mathbf{z}}\right)^2 - \left(k_{0n} a\right)^2}$, and $H^1_{(y,z)}$ the $\hat{\mathbf{y}}$ or $\hat{\mathbf{z}}$ component of the magnetic field. We can then derive a rigorous dispersion relation [31] for surface wave propagation:

$$D_n + F_n^+ + F_n^- = 0 \; , \; F_n^{\pm} = \overset{\infty}{\underset{m=1}{K}} \frac{-M^2/4}{D_{n \pm m}}, \tag{6}$$

whose convergence is guaranteed only for $v_m < v_{p,\min}$ where $v_{p,\min} = c\left(1 + 4(1+M)^2 \bar{X}_s^2\right)^{-1/2}$ is the value of $v_p$ associated with the maximal impedance $Z_{s,\max} = j\eta_0 \bar{X}_s (1+M)$. Beyond this value, the interaction between the modulation wave and the propagating wave may evoke instabilities and parametric amplification [32]. Due to the complicated nature of the continued fraction dispersion, we first take a qualitative, approximate look at its main properties [20,33] to provide physical insights into the effects of space-time modulation on the surface wave dispersion. Surface wave propagation over an isotropic impedance surface obeys a conical dispersion – the central orange cone in figure (2b). Space-time modulation replicates this cone along a diagonal line corresponding to the transformation $(k_z, \omega) \rightarrow (k_z + 2\pi/a, \omega + \Omega)$, whose slope is determined by $v_m$. An isofrequency cross-section is shown in figure (2c) for $\bar{X}_s = 4.5$, $a = 0.15\lambda_0$, modulation frequency $f_m = \Omega/2\pi = 0.155 f_0$, , yielding $v_p = c\left(1 + 4\bar{X}_s^2\right)^{-1/2} \approx 0.11c$, and $v_m \approx 0.028c$. The circles correspond to $\left(k_{tz}a + 2\pi n\right)^2 + \left(k_{ty}a\right)^2 = k_{0n}^2\left(1 + 4\bar{X}_s^2\right)$ (to reduce cluttering, not all harmonics are shown), and the smaller thick circles show the light cone cross sections. For small modulation depths $M$, the isofrequency dispersion of the modulated system differs from this schematic picture only close to the circle intersections, generating spatial bandgaps due to strong harmonic interaction near these points.

This is illustrated in figures (3a) and (3b), displaying the exact solution of the dispersion equation (6) without and with modulation, respectively, for $M \in [0.1, 0.4]$ (a surface wave solution must reside outside the light-cones shown in thick black lines). In the stationary case (Fig. (3a)), the dispersion is flat and symmetric with $k_z$, as expected, while in the synthetically moving scenario (Fig. (3b)), the dispersion shows a slope consistent with the approximate plot in Fig. (2c), and a one-way hyperbolic branch is obtained. The slope $s$ of the common tangent, which yields the hyperbolicity of the synthetically moving surface is given by

$$s = \left(v_p^2 / v_m^2 - 1\right)^{-1/2} = \frac{1}{\sqrt{\frac{a^2 f_m^2}{c^2}\left[1 + 4\bar{X}_s^2\right] - 1}}, \tag{7}$$

The bandgaps induced by coupling neighboring harmonics can be used to tailor the desired dispersion contour properties with large flexibility. For instance, in the canalization regime [1,10], of particular interest in the context of hyperbolic metsaurfaces to channel sub-diffractive details of an image to the far-field [34,35], all surface waves with a wide range of transverse momenta travel in the same direction, manifested as a straight line in the isofrequency contour. By tailoring the modulation signal, we can adjust the flatness of the dispersion, as seen in figures (3a),(3b). As the modulation depth increases, the outer curve flattens, and the internal, closed contours draw apart, realizing a response analogous to [1], corresponding to sub-diffraction imaging on the surface, additionally endowed with one-way nonreciprocal properties in figure (3b). As a result, large $k_y$ propagation is possible only towards certain directions, denoted by the tangent line. For small $M$, we can extend the perturbative approach in [36] to show that the width of the spectral gap created around the strongly coupled regions of the dispersion, where $n, n+1$ harmonics interact, is proportional to the modulation depth $M$:

$$\Delta k_{t,gap} a = \frac{\overline{X}_s^2 (k_{0,n} a)(k_{0,n+1} a)}{\sqrt{\mathbf{k}_{t,1} \cdot \mathbf{k}_{t,2}}} M , \qquad (8)$$

where $\mathbf{k}_{t,1}, \mathbf{k}_{t,2}$ are the wave vectors of the interacting surface waves corresponding to the $n, n+1$ (shown as thick black lines in figure (2c) [31]). The appearance of photonic bandgaps in time-modulated media is consistent with findings in [37].

*Excitation.* Hyperbolic propagation is of particular interest for the possibility of channeling high-momentum surface waves along specific directions. We examine this opportunity by studying a point source excitation of the proposed time-modulated metasurfaces. When time modulation is not present, coupling between all spatial harmonics supported by the surface occurs necessarily at the same frequency. Therefore, while coupling to the flat part of the dispersion using wave-vector matched excitation is possible [1,10], for instance through a carefully designed grating coupler, trying to excite a specific attribute of the dispersion using a localized emitter proves to be generally difficult, as most of the energy will couple to unwanted, typically low, wavenumbers. Figure (4a) shows the *x*-component of the electric field distribution when exciting a spatially modulated surface by a unit point current source ($I \cdot \ell = 1[A \cdot s]$ [38]). For moderate modulation depths ($M = 0.15$ here), the induced field profile looks almost isotropic, despite the complex band diagram associated with this surface (figure (3a)). Adding synthetic motion enables access not only enables one-way hyperbolic propagation, but interestingly also allows to isolate the extreme features of the dispersion diagram in figure (3b), as it yields a more complex system of coupling between spatial and temporal harmonics [18]. While the fundamental harmonic remains almost isotropic for excitation with a point source (figure (4b)), similar to the static scenario, it is now possible to excite higher-order harmonics through parametric mixing. These harmonics present pronounced non-reciprocal features and directional excitation of surface waves, as seen in figure (4c) for $n=3$, corresponding to $f = f_0 + 3f_m$, resembling the excitation of the moving metasurface shown in figure (1) [17], with large efficiency. The inset shows the electric field magnitude as a function of angle $\theta$, when sampled on a circle around the source with radii $R = 0.15\lambda_0$ (blue) and $R = 0.2\lambda_0$ (orange). The field is concentrated around the angle predicted by the slope of the dispersion in figure (3b), about 75° (schematically shown by the dashed purple line in figure (2c)), and it is emitted only towards the right of the source, due to the surface nonreciprocity.

We can use the same concept to couple the source to symmetric canalization of waves, using a standing wave modulation $Z_s = j\eta_0 \bar{X}_s \left(1 + M \cos[\kappa z] \cos[\Omega t]\right)$. In this case, figure (4d) shows again nearly-isotropic emission into the fundamental harmonic, while $n = 3$, presented in figure (4e), shows strong in-plane confinement and canalization. In Figure (4c), in order to synthesize one-way hyperbolic propagation towards a slanted angle, we were bound to relatively high modulation frequencies, since the slope $s$ increases with $f_m$ based on Eq. (7). Here we do not have this constraint, and can achieve canalization for smaller modulation frequencies, ideal for a photonic implementation. More examples of modulated and synthetically moving metasurfaces, and details regarding our full-wave simulations are provided in [31].

*Discussion*. While both mechanical and synthetic motions enable one-way hyperbolic response, important differences between the cases exist. Mechanical motion requires $v > v_p$ to support hyperbolic propagation, and the motion introduces *local* anisotropy to the surface or media properties (Eq. (1)) [16,17]. In the case of synthetic motion, hyperbolic propagation is supported in principle even for low values of $v_m$, and arbitrarily large modulation velocities may be achieved even with low modulation frequencies, making this approach attractive for practical purposes. However, when space-time modulation is considered, we introduce periodicity into the problem, which needs to be comparable to the wavelength of the surface wave on the unmodulated surface impedance (otherwise the circles in figure (2b) would be too far apart, and different harmonics would not interact). The effect is thus inherently nonlocal, and it is associated with asymmetric coupling of adjacent harmonics. Interestingly, the hyperbolic regime exists for complementary ranges of velocities ($v, v_m$): for a moving surface the velocity must be greater than the phase velocity of the surface waves at rest, while in the modulated case the modulation velocity must be smaller than the phase velocity $v_{p,\min}$ [31] to avoid instabilities (although stable regimes for higher velocity exist, but they require fast modulation frequencies). We can define an effective velocity $v_{eff}$ induced by the synthetic modulation velocity

$$v_{eff} = v_p \sqrt{\frac{1}{1 - 4\bar{X}_s^2 (v_m / c)^2}} , \qquad (9)$$

defined as the mechanical velocity that provides the same slope of the hyperbolic dispersion. This expression, combined with the limitation $v_m < v_{p,\min}$ reveals a relevant tradeoff of synthetically moving hyperbolic metasurfaces: on one hand, larger $M$ enables stronger coupling to higher-order harmonics that sustain stronger excitation of one-way plasmons, but at the same time $v_{p,\min}$ decreases, limiting the effective synthetic velocity $v_{eff}$, as further elaborated in [31]. In order to shed light on the directional nature of propagation over the proposed metasurface (shown for $n=3$ in figure (4)), in [31] we calculate $H_n / H_0$ along the dispersion curves, showing that higher harmonics are more dominant in the flatter parts of the dispersion and enabling efficient frequency conversion with directional and non-reciprocal properties [31].

In summary, we have shown that space-time modulation of a surface impedance provides a flexible platform to mimic mechanical motion, enabling access to intriguing dispersion features such as extreme anisotropy, flat dispersion, canalization and non-reciprocal hyperbolic propagation. The required traveling-wave modulation may be imparted by acoustic or mechanical waves traveling over metasurfaces or 2D materials. These findings may open interesting opportunities in the realization of non-reciprocal sub-diffractive imaging systems, highly unusual plasmon transport structures, and enhanced non-reciprocal light-matter interactions based on the proposed platform.

*Acknowledgements*. We would like to thank Dr. Dimitrios Sounas for assistance with full-wave electromagnetic modeling.

**References**

*To whom correspondence should be addressed: aalu@gc.cuny.edu

**Figures**

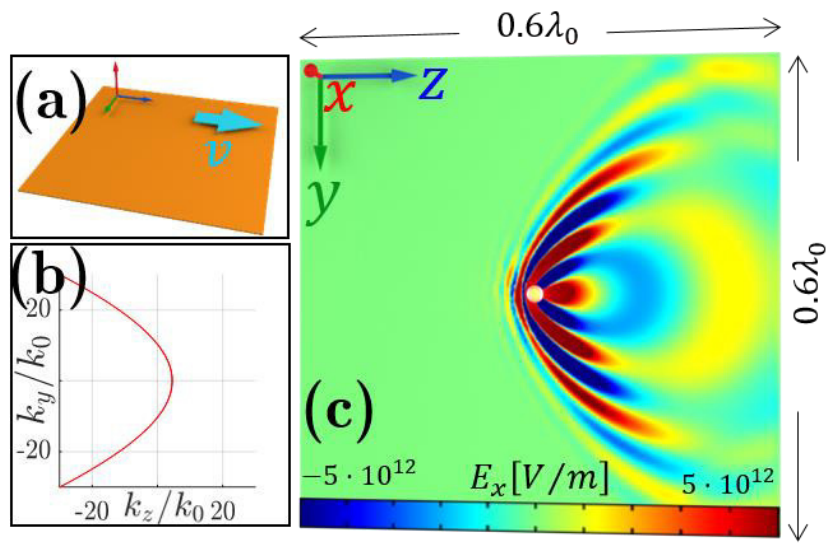

**Figure 1.** (a) The geometry of the system. (b) An iso-frequency dispersion curve for $v = 0.114c$, which is above the threshold value. (c) Nonreciprocal hyperbolic response of a moving impedance surface, excited by a point source.

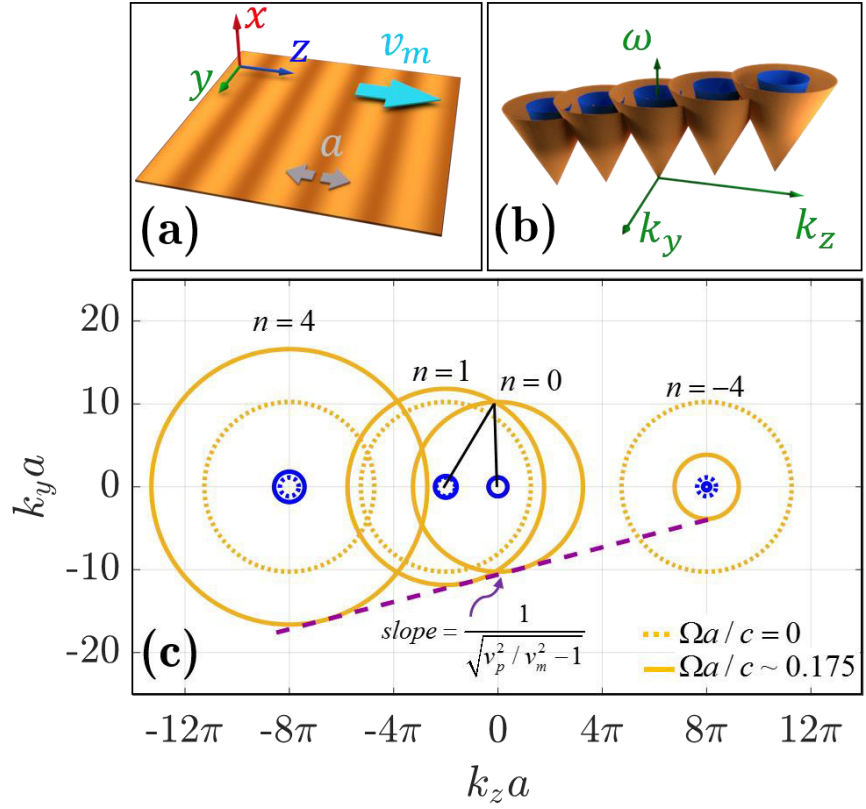

**Figure 2.** (a) A schematic of the system. A periodically modulated impedance surface, with spatial period $a$, modulation wave velocity $v_m$. (b) Weak space-time modulation replicates the basic dispersion cone along the transformation $(k_z, \omega) \to (k_z + 2\pi/a, \omega + \Omega)$. The modulation couples the cone replicas, altering the actual dispersion only close to their intersections, which creates spatial and temporal bandgaps. (c) An isofrequency cross-section of (b). The stationary scenario is shown in dotted for comparison. The purple dashed line is the common tangent.

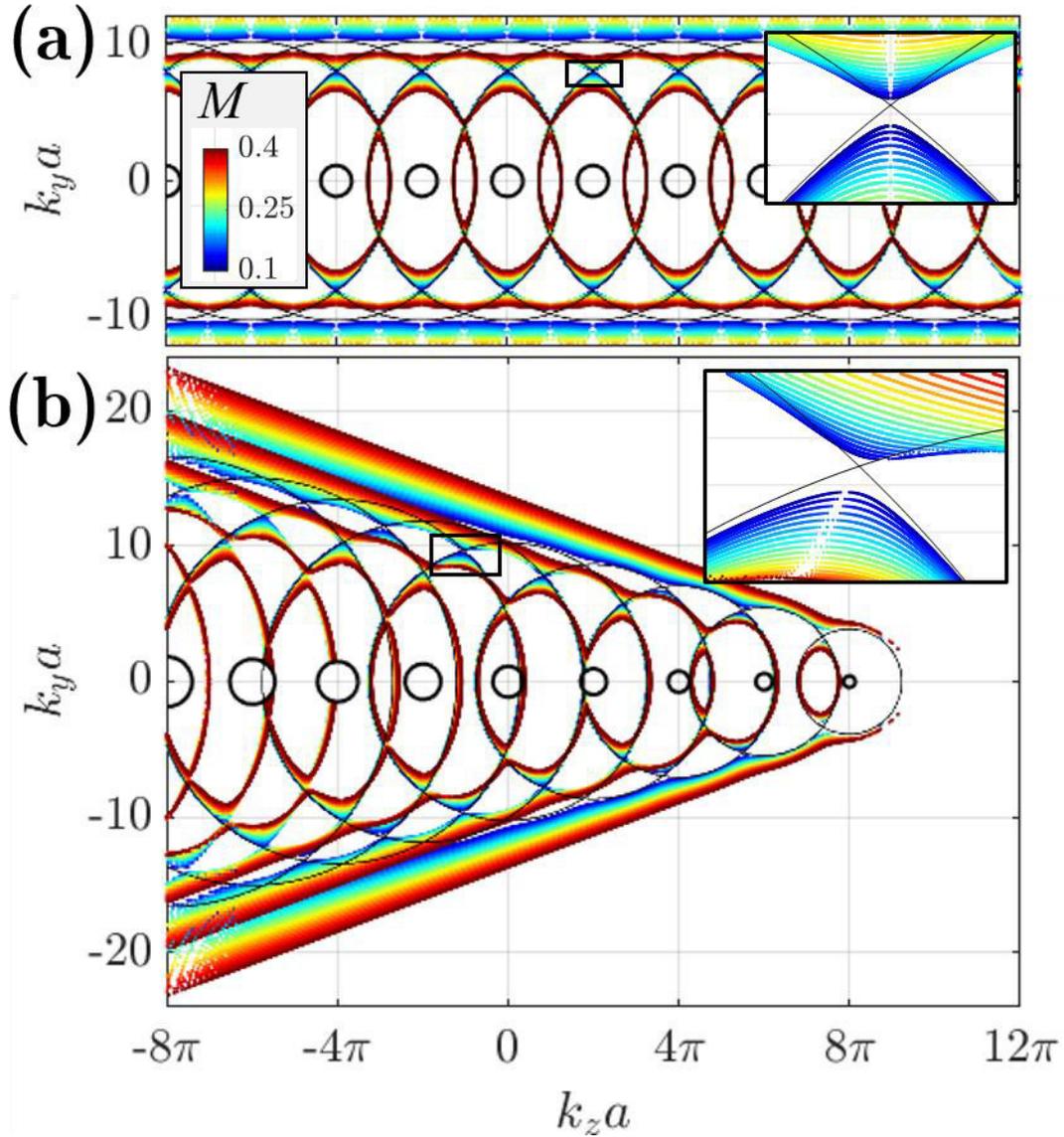

**Figure 3**. (a) An isofrequency dispersion contour derived from the exact dispersion relation in equation (6), without time modulation. The system is $z$-periodic in space, and hence also $k_z$ periodic, with period $2\pi/a$. Inset shows bandgap widening as $M$ increases. (b) Same as (a) but with travelling wave space-time modulation. The diagonal dispersion lines are formed, associated with propagation of higher-order harmonics.

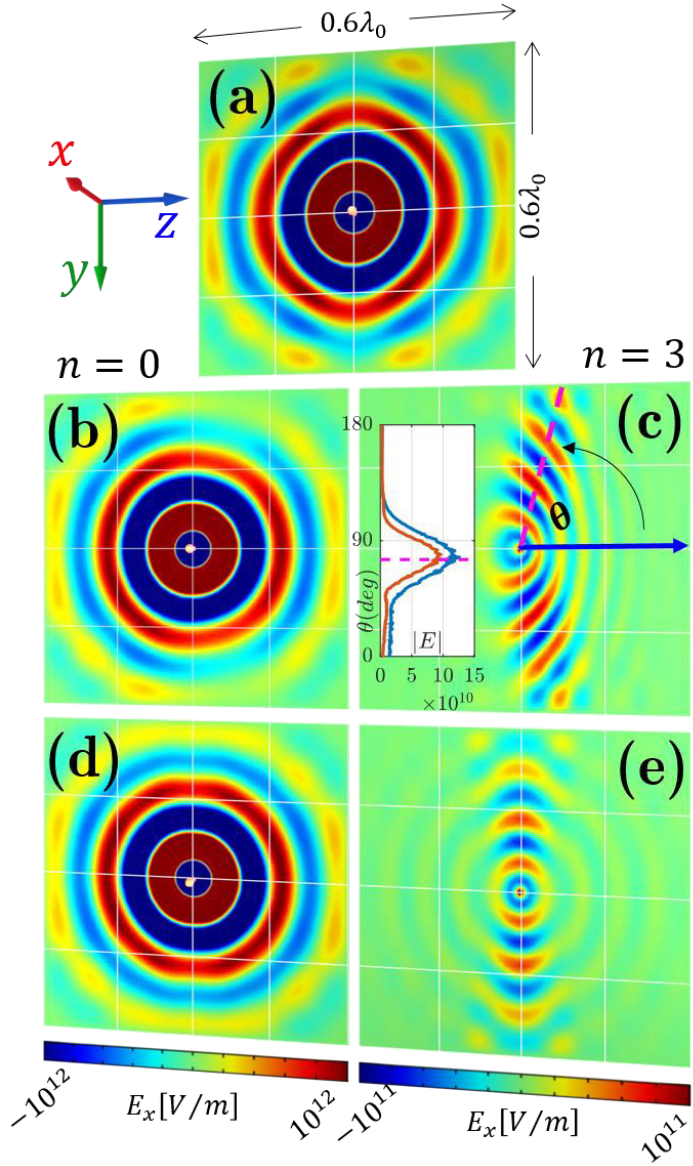

**Figure 4**. Excitation of a space-time modulated metasurface by a point source. For all cases $M = 0.15$ and $a \approx 0.18\lambda_0$. The source is shown in the center, and the distance between neighboring grid lines is $0.15\lambda_0$. (a) Only space-modulation is present, corresponding to $\Omega = 0$. A nearly isotropic response is observed. (b) Fundamental harmonic when travelling wave modulation is applied. The hyperbolic dispersion is weakly excited by a localized point source. (c) $n = 3$ harmonic. Non-reciprocal hyperbolic propagation is evident, as collimated beams

propagate towards specific directions, consistent with the mechanical motion considered in Fig. 1. The inset shows the electric field magnitude sampled on a circular arc around the source with radius $R = 0.15\lambda_0$ in orange, and $0.2\lambda_0$ in blue. (d) $n = 0$ for standing-wave modulation. (e) The point source couples strongly to a narrow propagation direction for the $n = 3$ harmonic, corresponding to the predicted flat dispersion.

# One-Way Hyperbolic Metasurfaces Based on Synthetic Motion – Supporting material

Yarden Mazor and Andrea Alù[*]

## I. Representation of space-time modulation using modulated capacitors and inductors

If the modulated surface impedance is implemented using a surface distribution of inductors and capacitors, the boundary condition can be expressed using the following operator [1]

$$\mathbf{E}_{\tan}(z,t) = \hat{Z}_s\left[\mathbf{J}_s(z,t)\right] = L_s(z,t)\frac{\partial \mathbf{J}_s(z,t)}{\partial t} + C_s^i(z,t)\int^t \mathbf{J}_s(z,t) \tag{S1}$$

Since we are interested in the case of both $L_s$ and $C_s$ corresponding to a travelling wave modulation of the surface parameters, they can both be written as

$$L_s(z,t) = \sum_{n'=-\infty}^{\infty} L_{n'} e^{-jn'\kappa z} e^{jn'\Omega t} \quad , \quad \frac{1}{C_s} = \sum_{n'=-\infty}^{\infty} C_{n'}^i e^{-jn'\kappa z} e^{jn'\Omega t} \tag{S2}$$

and since both are real quantities, the coefficients must satisfy

$$L_{-n'} = L_{n'}^* \quad , \quad C_{-n'}^i = C_{n'}^{i*}. \tag{S3}$$

If we substitute this representation into the impedance boundary condition $\mathbf{E}_{\tan} = \hat{Z}_s\left[\hat{\mathbf{n}} \times (\mathbf{H}_2 - \mathbf{H}_1)\right]$ and use the field representation

$$\mathbf{H} = \sum_{n=-\infty}^{\infty} H_n^{1,2}\left(\cos\varphi_n \hat{\mathbf{z}} - \sin\varphi_n \hat{\mathbf{y}}\right) e^{\mp \alpha_n x - j k_{ty} y - j(k_{tz} + n\kappa)z} e^{j\omega_n t} + C.C \tag{S4}$$

As shown in equation **(4)** of the article, we obtain

$$\sum_{n=-\infty}^{\infty}\left(j\omega_n L_{m-n}+\frac{1}{j\omega_n}C^i_{m-n}+\frac{j\alpha_n}{2\varepsilon_0\omega_n}\delta_{m-n}\right)H^1_{n,(y,z)} \ , \ \forall m \tag{S5}$$

which yields an infinite set of linear equations. Using this boundary condition instead of the one in equation **(3)** of the article, takes into account the temporal dispersion of the surface impedance.

## II.  2D dispersion for space-time modulated surface and spatial bandgap

If we substitute the field expansion (equation (S4)) into the appropriate boundary condition (either (S1) or **(5)+(3)** of the article with the assumption of cosine travelling wave modulation instead of a general periodic function), we obtain the equation governing the coupling between the different harmonics

$$Z_{n-1}H^1_{n-1,(y,z)}+\left(Z_n+\frac{j\alpha_n}{2\varepsilon_0\omega_n}\right)H^1_{n,(y,z)}+Z_{n+1}H^1_{n+1,(y,z)}=0 \tag{S6}$$

Which assumes the travelling modulation is a harmonic wave of the form $1-M\cos(\kappa z-\Omega t)$. $H^1_{n,(y,z)}$ refers to $y,z$ components of the magnetic field. For a more general modulation scheme, (S6) will include additional harmonics. In this case, equation (S2) has only 3 terms

$$L_0=L \ , \ L_{\pm 1}=\frac{ML}{2} \ , \ C^i_0=C^i \ , \ C^i_{\pm 1}=\frac{MC^i}{2} \tag{S7}$$

and we obtain equation (S6) with

$$Z_n = j\omega_n L_{m-n} + \frac{1}{j\omega_n} C^i_{m-n} \tag{S8}$$

for any chosen value of $m$. The $Z_n$ coefficients generally depend on frequency due to the temporal dispersion of the surface parameters, discussed in the previous section. If all parameters are modulated with the same $M, \Omega, a$, and we operate in regions where the impedance can be considered relatively non-dispersive (e.g., far from resonance, $\Omega \ll \omega$, or any other mechanism that would help keep the impedance constant in the relevant frequency range), we can neglect the dependence of $Z_n$ on frequency for the dominant harmonics, and use equation **(3)** of the article, leading to a simpler solution and providing physical insights into the problem.

$$H^1_{n-1,(y,z)} + \frac{2}{M} D_n H^1_{n,(y,z)} + H^1_{n+1,(y,z)} = 0 \tag{S9}$$

with

$$D_n = 1 - \frac{\sqrt{(k_y a)^2 + (k_z a + 2n\pi)^2 - (k_{0n} a)^2}}{2\bar{X}_s k_{0n} a} = 1 - \frac{\sqrt{(\mathbf{k}_t a + 2\pi n \hat{\mathbf{z}})^2 - (k_{0n} a)^2}}{2\bar{X}_s k_{0n} a}. \tag{S10}$$

The solutions to the equation $D_n(k_y, k_z) = 0$ yield the circles shown in figure **(2c)** in the article. Following the same steps shown in [2–4], we arrive at a dispersion equation represented using continued fractions

$$D_n - \cfrac{M^2/4}{D_{n-1} - \cfrac{M^2/4}{D_{n-2} - \cfrac{M^2/4}{D_{n-3} - \ldots}}} - \cfrac{M^2/4}{D_{n+1} - \cfrac{M^2/4}{D_{n+2} - \cfrac{M^2/4}{D_{n+3} - \ldots}}} = 0 \tag{S11}$$

And the solutions $(k_y, k_z)$ to this equation yields the dispersion curve. Throughout this paper, the continued fractions were evaluated using the modified Lentz algorithm [5,6]. A compact notation uses the symbol $K$, in the following form

$$b_0 + \cfrac{a_1}{b_1 + \cfrac{a_2}{b_2 + \cfrac{a_3}{b_3 + ...}}} = b_0 + \underset{n=1}{\overset{\infty}{K}} \frac{a_n}{b_n} \tag{S12}$$

Which lets us rewrite (S11) as

$$D_n + \underset{m=1}{\overset{\infty}{K}} \frac{-M^2/4}{D_{n+m}} + \underset{m=1}{\overset{\infty}{K}} \frac{-M^2/4}{D_{n-m}} = 0 \tag{S13}$$

For the continued fraction to converge, we must satisfy $\left|\frac{2}{M} D_n\right| > 2$, $\forall |n| > n_0$ for some finite $n_0$ (equivalent to the requirement $\lim_{|n| \to \infty} D_n > 2$). This yields the constraint

$$v_m = \frac{\Omega a}{2\pi} < \frac{c}{\sqrt{1 + 4(1+M)^2 \bar{X}_s^2}} \tag{S14}$$

meaning that the modulation wave velocity must be smaller than the local TM wave velocity guided on any point of the surface. For $M \ll 1$ we can get a perturbative solution for the dispersion by using the following equivalent representation to equation (S11)

$$\left( D_n - \cfrac{M^2/4}{D_{n-1} - \cfrac{M^2/4}{D_{n-2} - \cfrac{M^2/4}{D_{n-3} - ...}}} \right) \left( D_{n+1} - \cfrac{M^2/4}{D_{n+2} - \cfrac{M^2/4}{D_{n+3} - ...}} \right) = \frac{M^2}{4} \tag{S15}$$

and following the considerations presented in [7] we may represent the dispersion around the intersection point approximately as

$$D_n D_{n+1} = \frac{M^2}{4} \tag{S16}$$

Now, let us denote the in-plane wave-vector corresponding to the intersection of these circles as $\mathbf{k}_{t,s}$, and represent the solution to equation (S16) as $\mathbf{k}_t = \mathbf{k}_{t,s} + \mathbf{k}'$ with $|\mathbf{k}'| \ll |\mathbf{k}_{t,s} + 2n\pi\hat{\mathbf{z}}|$. It is convenient to use the vectors shown in figure **(2c)** in the article using black lines. These vectors are $\mathbf{k}_{t,1} = \mathbf{k}_{t,s} + 2n\pi\hat{\mathbf{z}}$, $\mathbf{k}_{t,2} = \mathbf{k}_{t,s} + 2(n+1)\pi\hat{\mathbf{z}}$. If we substitute these assumptions into equation (S16) we obtain

$$\left( \mathbf{k}_{t,1} \cdot \mathbf{k}' \right) \left( \mathbf{k}_{t,2} \cdot \mathbf{k}' \right) = 4M^2 \bar{X}_s^4 \left( k_{0,n} a \right)^2 \left( k_{0,n+1} a \right)^2 \tag{S17}$$

This equation represents a hyperbola which approximates the dispersion around the spatial bandgaps. By using geometric properties of this hyperbola [8], we may extract its major axis – being an accurate measure of the spatial bandgap width shown in equation (9) of the article, for small modulation depths.

# III. Numerical modeling and convergence of space-time modulated surfaces in COMSOL Multiphysics

Assume we have a surface with an impedance that is a periodic function of time, and may include also dependence on the spatial coordinates. It can be represented by its temporal Fourier series

$$\sigma = \sum_{n=-\infty}^{\infty} \sigma_n(y,z) e^{-jn\Omega t} \ . \tag{S18}$$

The electromagnetic fields may also be represented in the same manner

$$\begin{aligned}\mathbf{E}(\mathbf{r},t) &= \sum_{n=-\infty}^{\infty} \mathbf{E}_n(\mathbf{r}) e^{j\omega t + jn\Omega t} \\ \mathbf{H}(\mathbf{r},t) &= \sum_{n=-\infty}^{\infty} \mathbf{H}_n(\mathbf{r}) e^{j\omega t + jn\Omega t}\end{aligned} \ . \tag{S19}$$

If the modulation is in the form presented in the main text, including only cosine dependence on time, the Fourier series in (S18) can be written using 3 terms only, for example

$$\sigma(y,z,t) = \sigma\left(1 + M\cos[\kappa z - \Omega t]\right) \Rightarrow \sigma_0 = \sigma \ , \ \sigma_1 = \frac{1}{2} M\sigma e^{-j\kappa z} \ , \ \sigma_{-1} = \frac{1}{2} M\sigma e^{j\kappa z} \quad (S20)$$

We substitute the fields and the impedance expression into the impedance boundary condition, and obtain

$$\hat{\mathbf{x}} \times \left( \sum_{n=-\infty}^{\infty} \mathbf{H}_{n,x=0^+}(\mathbf{r}) e^{j\omega t + jn\Omega t} - \sum_{n=-\infty}^{\infty} \mathbf{H}_{n,x=0^-}(\mathbf{r}) e^{j\omega t + jn\Omega t} \right) = \left( \sigma + \sigma_1 e^{j\Omega t} + \sigma_{-1} e^{-j\Omega t} \right) \sum_{n=-\infty}^{\infty} \mathbf{E}_{n,\tan}(\mathbf{r}) e^{j\omega t + jn\Omega t}$$

(S21)

which can be rearranged into

$$\sum_{n=-\infty}^{\infty} \hat{\mathbf{n}} \times \left( \mathbf{H}_{n,x=0^+} - \mathbf{H}_{n,x=0^-} \right) e^{j(\omega+n\Omega)t} = \sum_{n=-\infty}^{\infty} \left( \sigma_0 \mathbf{E}_{n,\tan} + \sigma_1 \mathbf{E}_{n-1,\tan} + \sigma_{-1} \mathbf{E}_{n+1,\tan} \right) e^{j(\omega+n\Omega)t} = \sum_{n=-\infty}^{\infty} \mathbf{J}_{s,n} e^{j(\omega+n\Omega)t}$$

(S22)

This equation can be used to model this system in full-wave simulations using the following scheme:

- The solution is truncated to a finite number of harmonics. The validity of this step can be later checked by adding additional harmonics and verifying that the solution in the frequency of interest is not significantly altered. We will refer to the number of harmonics in the truncated solution as $N$. Naturally, the amount of variables to be solved by the simulation software is associated to the value of $N$, which is limited by the computational resources. In our case, we could go up to $N=5$ and maintain reasonable calculation times, and therefore this value is regarded as the reference for error calculations.

- Each harmonic $n$, corresponding to frequency $f_0 + nf_m$ (where $f_m = \Omega/2\pi$) is assigned an electromagnetic wave solver in the solution domain.

- The point source (or any other source of interest) is placed in the instance corresponding to the fundamental frequency $f_0$.

- The surface current for each harmonic is defined by equation (S22) in the "surface current density" boundary condition, using the electric field expressions from the relevant instances.

To examine the convergence of the numerical full-wave solution let us look at the fields obtained for the $n=0$ and $n=3$ harmonic for various values of $N$, as seen in figure **(1)**.

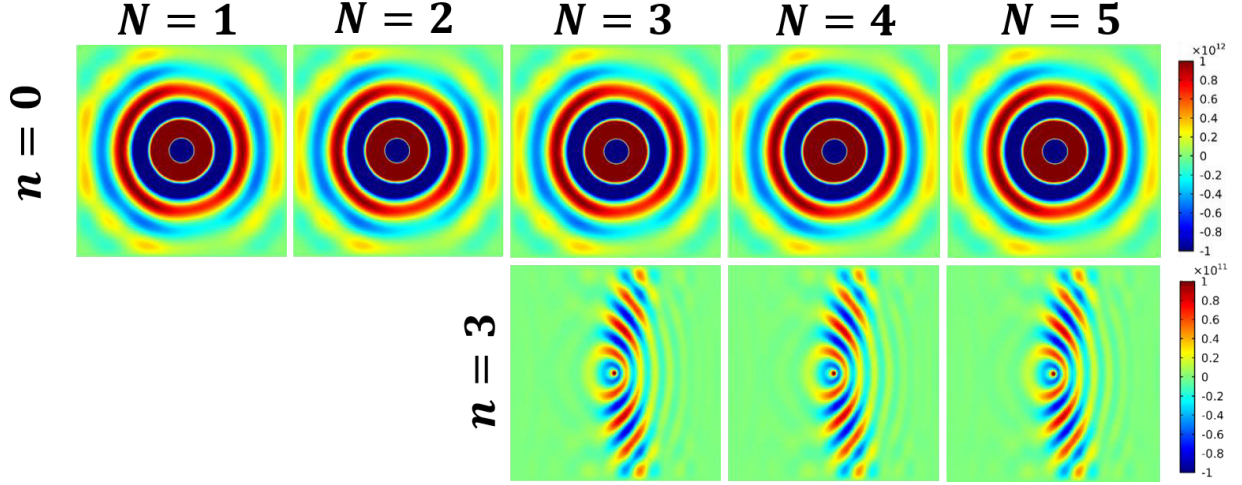

*Figure 1. The x-component of the electric field for the $n=0$ and $n=3$ harmonics for various truncations, denoted by the value of $N$. We see that the solution does not change significantly when adding additional harmonics, implying its convergence.*

We see that there are almost no visible changes when adding more harmonics to the truncation, indicating convergence. To get a more quantitative look at this, we examine the electric field norm along a circular arc around the source location, with radius of $100 \mu m$, shown in figure **(2)**.

We can define the relative error for a specific truncation $N$, and for a field harmonic $n$ (corresponding to the harmonics in the expansion in equation (S19)) as

$$\varepsilon_n(N) = \frac{\left\| |\mathbf{E}_n(N)| - |\mathbf{E}_n(N_{ref})| \right\|}{|\mathbf{E}_n(N_{ref})|} \tag{S23}$$

Where $\mathbf{E}_n(N)$ is the n'th harmonic of the electric field for a solution truncated after $N$ terms, and $N_{ref}$ is chosen here as $N_{ref} = 5$. The errors are shown in figure (3).

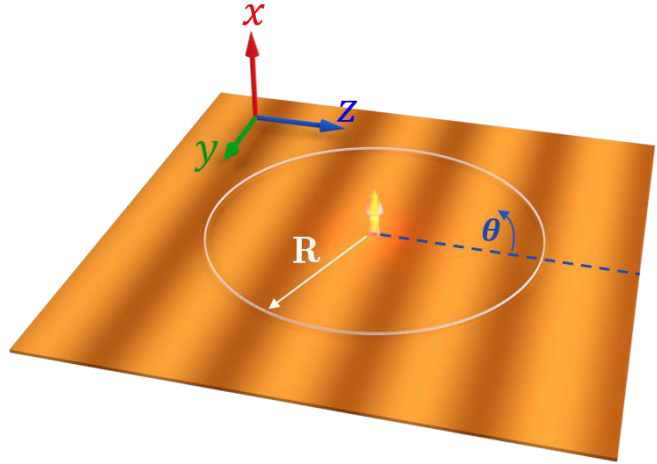

*Figure 2. The circular arc used to sample the electric field for the calculation of the relative errors shown in figure (3).*

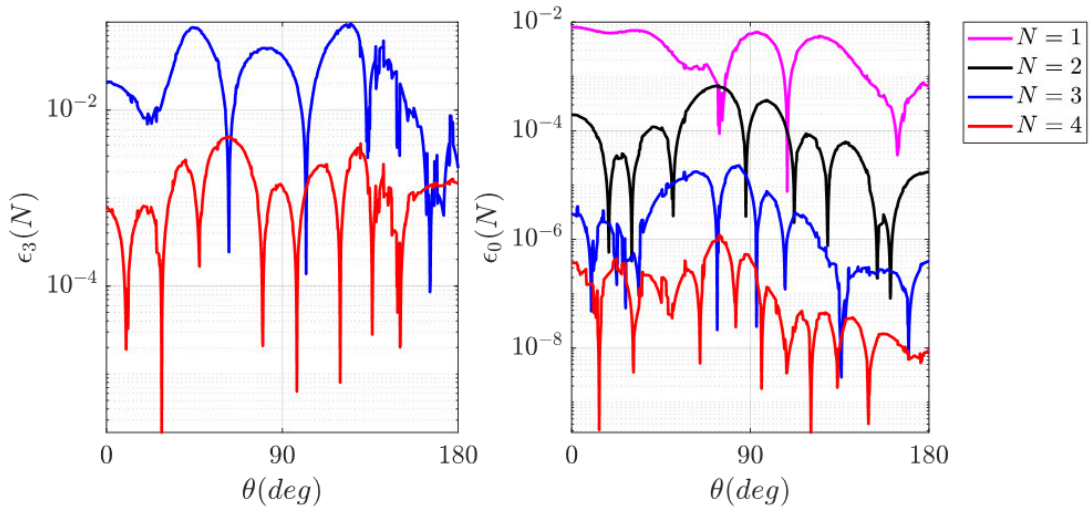

*Figure 3. Relative errors of the $n = 0$ and $n = 3$ with various truncations, with respect to the truncation $N = 5$. We see that as we take more and more harmonics, the error decreases.*

## IV. Canalization for smaller modulation frequencies

As mentioned in the article, if we are interested in achieving canalization of waves in a symmetric way, without incorporating one-way hyperbolic propagation, we can use smaller modulation frequencies than the ones shown in the article. Figure **(4)** shows the obtained response for $f_m = 10[GHz]$ (in the article $70[GHz]$ is used) in the same way presented in figure **(4)** in the main article. Using lower modulation frequency renders the directional response at a certain frequency much closer to the base harmonic

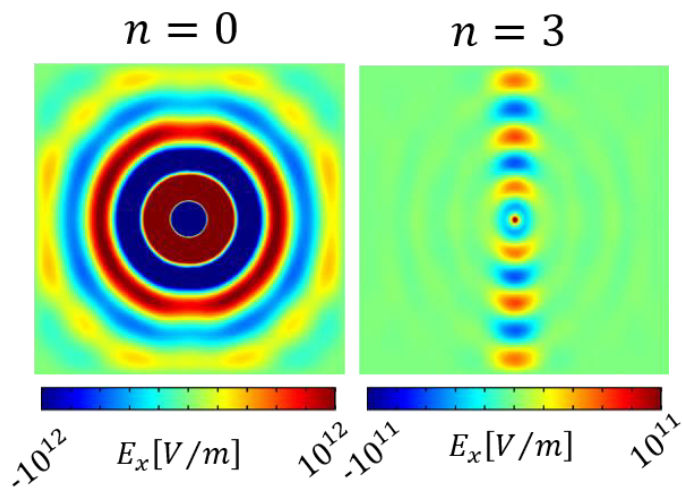

*Figure 4. Response of the $n=0$ and $n=3$ harmonics for standing wave modulation, with frequency $f_m = 10[GHz]$. Strong anisotropy is visible.*

## V. Tradeoffs concerning the frequency conversion efficiency and effective velocity

Figure (5a) shows the $n=3$ field intensities as a function of $M$, and in the inset $v_{eff}$ as a function of $v_m$, with maximum possible $v_m$ depicted as a colored vertical line corresponding to a given value of $M$. The intersection of these vertical lines with the $v_{eff}(v_m)$ curve indicates the maximum $v_{eff}$ that can be synthesized for a given value of $M$. As mentioned in the main article, we see that as $M$ increases we get a more efficient upconversion, yet the equivalent mechanical speeds we are able to mimic become smaller. In figure (5b) we see the results of calculating the ratio $H_n/H_0$ along the dispersion isofrequency curve, for $M = 0.15$, where we display using the color coding the most dominant harmonic ($\max_n\{H_n/H_0\}$). We can see that the higher harmonics are the dominant ones in the flatter regions of the dispersion. As an example, the $n = 3$ harmonic becomes dominant in the outer regions of the dispersion (shown with larger markers). Hence, we can expect that a confined source exciting a broad range of spatial harmonics, such as the point source presented in the main article, couples strongly to highly directional waves for the upconverted harmonics – as shown in figure (4) of the main article. For comparison, in blue we show the dispersion presented in figure (1b) of the main article, corresponding to $v_{eff}$ for the chosen parameters, highlighting the strong analogy between the two systems in terms of overall response.

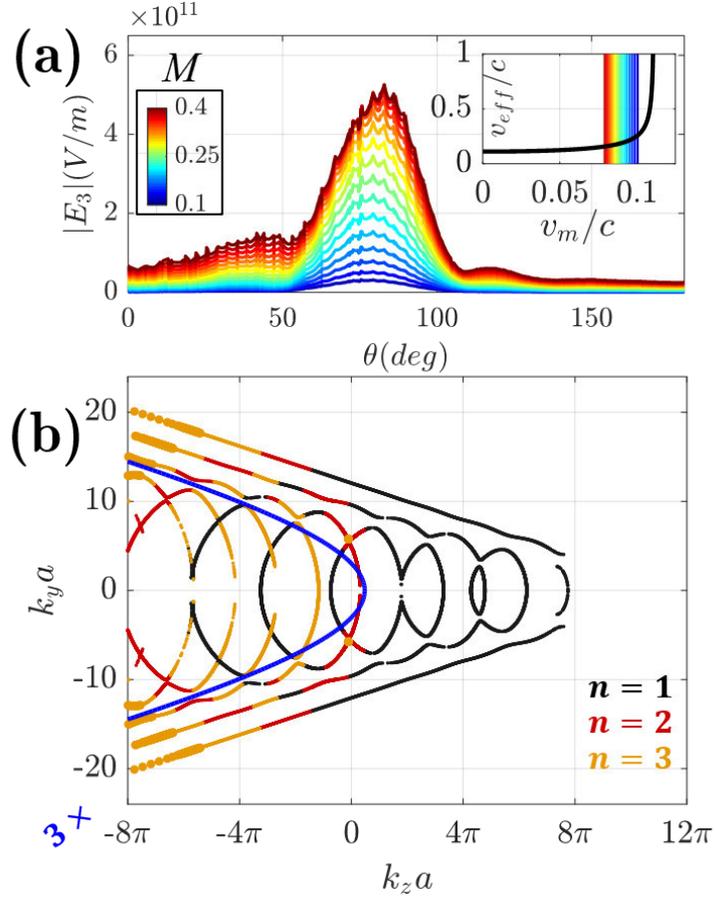

*Figure 5. (a) Electric field amplitude sampled on a circle around the source, for the third-order harmonic vs M. The inset shows the modulation velocity required to generate a dispersion slope similar to a moving homogeneous surface with velocity $v_{eff}$. (inset) Maximum $v_m$ with the same color code, demonstrating the discussed tradeoff. (b) A map of the most dominant harmonics for $M = 0.15$ (large markers indicate more pronounced contributions). In blue we show the hyperbolic dispersion of a mechanically moving homogeneous surface with $v_{eff} = 0.114c$. The axes for the blue line overlay are multiplied by a factor of 3, which preserves the slope.*